\documentclass[aps,prl,twocolumn,superscriptaddress,floatfix,nofootinbib]{revtex4-2}

\usepackage{graphicx}
\usepackage{amsmath}
\usepackage{amssymb}
\usepackage{hyperref}
\usepackage{xcolor}
\usepackage{booktabs}
\usepackage{siunitx}
\graphicspath{{./}}

\begin{document}

\title{Beyond Beryllium: AI-Accelerated Materials Discovery
for Interstellar Spacecraft Shielding}

\author{Yue Li$^{*}$}
\affiliation{School of Materials Science and Engineering,
Nanyang Interstellar University, Singapore 639798, Singapore}

\author{Xu Pan$^{*}$}
\affiliation{School of Artificial Intelligence,
Nanyang Interstellar University, Singapore 639798, Singapore}

\author{Kaiyuan Guo$^{*}$}
\affiliation{Department of Medical Radiation Management and Hibernation,
Shanghai Interstellar Jiao Tong University, Shanghai 200240, China}

\date{April 1, 2026}

\begin{abstract}
Project Daedalus (1973--1978), the most detailed interstellar probe
design study ever conducted, specified a 9~mm beryllium erosion shield
to protect the spacecraft payload during its 5.9~light-year cruise to
Barnard's Star at 12\% of the speed of light. This design, however,
predated both the isolation of two-dimensional materials and the
development of graph neural network (GNN) property predictors.
Here, we systematically screen 20 candidate materials---spanning
conventional aerospace metals, transition metal dichalcogenides,
and ultra-high-temperature ceramics---using density functional theory
(DFT) data from the JARVIS database (76{,}000 materials) with
independent validation by the Atomistic Line Graph Neural Network
(ALIGNN). We evaluate candidates across four criteria: specific
mechanical stiffness ($K_V/\rho$), sputtering resistance, thermal
neutron absorption cross-section, and thermodynamic stability.
Our screening identifies hexagonal boron nitride (h-BN) and boron
carbide (B$_4$C) as dual-function materials offering simultaneous
mechanical protection and neutron radiation shielding, and we propose
a graphene/h-BN/polymer layered heterostructure shield design that
achieves an estimated 47\% mass reduction relative to the original
beryllium specification. These findings will become immediately
actionable upon the successful development of fusion pulse propulsion,
which we note remains an outstanding engineering challenge.
\end{abstract}

\maketitle

% Equal contribution footnote on page 1
\renewcommand{\thefootnote}{*}
\footnotetext{These authors contributed equally and all serve as
corresponding authors. E-mail: liyue2341@gmail.com (Y.L.),
kimpan2341@gmail.com (X.P.), Guoky14@gmail.com (K.G.)}
\renewcommand{\thefootnote}{\arabic{footnote}}

% TOC graphic (unnumbered)
\begin{center}
\includegraphics[width=0.92\columnwidth]{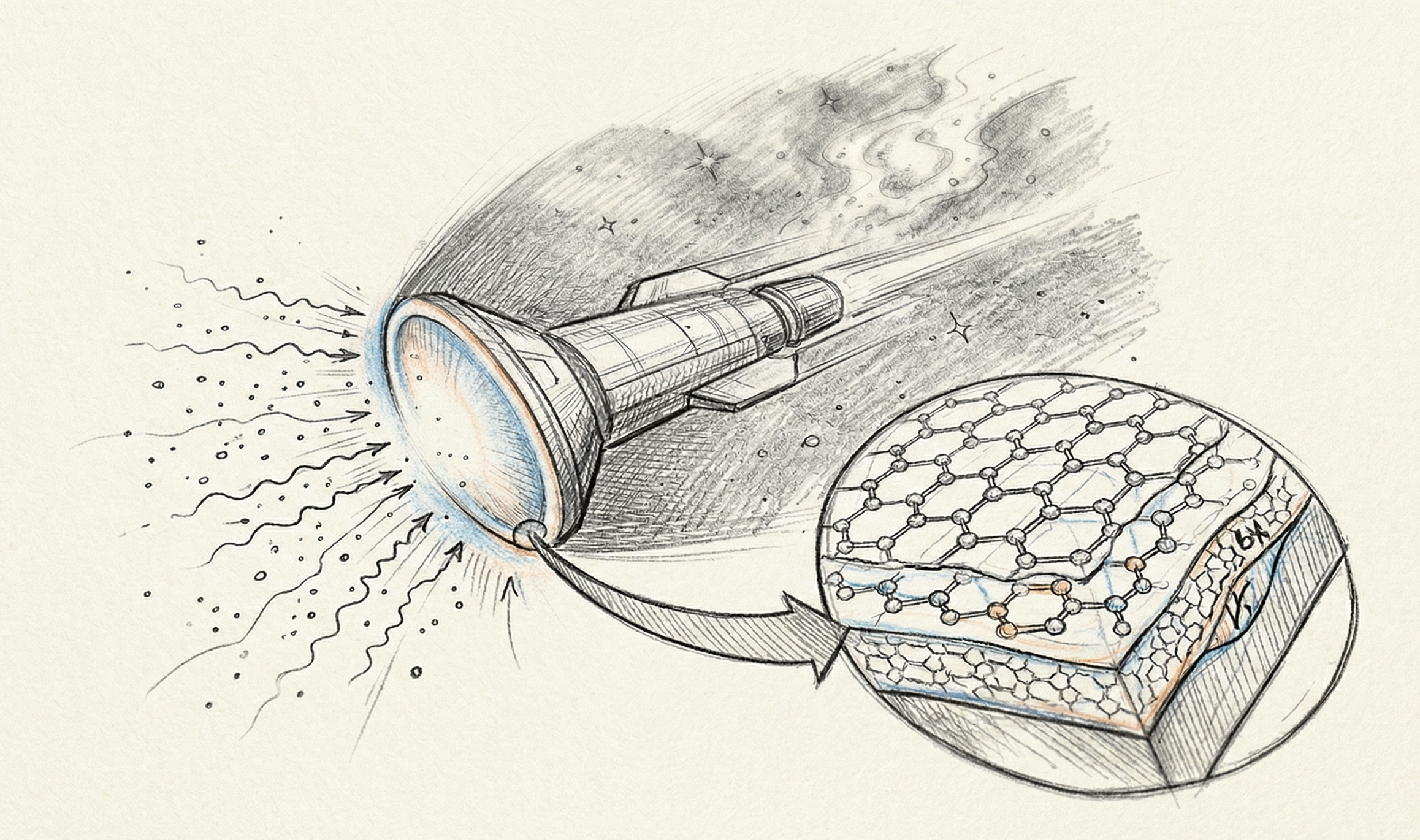}\\[2pt]
{\small\textit{Table of Contents Graphic}:
An interstellar probe encounters the ISM at $0.12c$.
Inset: layered heterostructure shield with graphene and
h-BN nanostructures.}
\end{center}
\vspace{4pt}

%=====================================================================
\section{Introduction}
%=====================================================================

The prospect of interstellar travel has motivated some of the most
ambitious engineering studies in human history. In 1968,
Dyson~\cite{Dyson1968} articulated the fundamental energetics of
interstellar transport, establishing that nuclear pulse propulsion
could in principle accelerate a spacecraft to a significant fraction
of the speed of light. Project Orion (1958--1965) explored nuclear
pulse propulsion using fission bombs before being curtailed by the
Partial Test Ban Treaty~\cite{Schmidt2000}. The most comprehensive
design study to date, Project Daedalus (1973--1978), proposed an
unmanned probe propelled by inertial confinement fusion to reach
Barnard's Star, 5.9~light-years distant, at a cruise velocity of
$0.12c$~\cite{Bond1978}. More recently, Project Icarus has sought
to update the Daedalus concept with modern engineering
knowledge~\cite{Long2009}.

A critical and often underappreciated challenge of relativistic
spaceflight is the bombardment of the spacecraft by interstellar
medium (ISM) particles. At $0.12c$ ($3.6 \times 10^7$~m/s), even
the tenuous ISM becomes a formidable particle beam. Martin's analysis
for the Daedalus study~\cite{Martin1978} showed that over a
5.9~light-year transit through a region with particle density
$n \approx 1$~cm$^{-3}$, the frontal shield (area $A \approx
491$~m$^2$) would encounter $\sim 2.7 \times 10^{25}$ particles,
with individual protons carrying kinetic energies of
$\sim$6.7~MeV---well into the nuclear reaction regime. Larger dust
grains, though far rarer, can deliver megajoule-scale impacts
equivalent to macroscopic explosions~\cite{Long2023,Hoang2017}.

The Daedalus team's solution was a 9~mm beryllium erosion shield,
selected for its combination of low density
($\rho = 1.85$~g/cm$^3$), reasonable bulk modulus
($K_V \approx 115$~GPa), and high sublimation energy
($E_{\rm sb} = 3.36$~eV/atom)~\cite{Bond1978,Martin1978}. However,
this design was constrained to the materials knowledge of the 1970s.
In the intervening half-century, materials science has undergone
transformative advances: the isolation and characterization of
two-dimensional materials beginning with graphene in
2004~\cite{Lee2008graphene}; the development of ultra-high-temperature
ceramics (UHTCs) with extreme mechanical
properties~\cite{Fahrenholtz2007uhtc}; and the discovery that
boron-containing materials provide exceptional neutron radiation
shielding~\cite{Thibeault2012,Harrison2008,BNNP2024foam,Kim2025bnnt}.

Equally transformative has been the rise of computational materials
screening. The JARVIS (Joint Automated Repository for Various
Integrated Simulations) database now contains density functional
theory (DFT) calculations for over 76{,}000
materials~\cite{Choudhary2020jarvis}, while graph neural network
architectures such as the Atomistic Line Graph Neural Network
(ALIGNN) enable rapid property prediction with near-DFT
accuracy~\cite{Choudhary2021alignn,Choudhary2023phonon,Choudhary2018elastic}.

In this work, we leverage the JARVIS-DFT database and ALIGNN
pretrained models to systematically re-evaluate the materials
selection for interstellar dust shielding. We screen 20 candidate
materials across three families---conventional aerospace metals,
layered/two-dimensional materials, and ceramics/superhard
compounds---against four performance metrics relevant to the Daedalus
mission profile. We identify several materials that substantially
outperform beryllium, propose a layered heterostructure shield
concept, and discuss the implications for future interstellar
mission design.

%=====================================================================
\section{Methods}
%=====================================================================

\subsection{Mission Parameters}

We adopt the Daedalus Phase~2 mission profile~\cite{Bond1978}:
cruise velocity $v = 0.12c = 3.6 \times 10^7$~m/s, distance to
Barnard's Star $d = 5.9$~ly $= 5.58 \times 10^{16}$~m, yielding
a cruise time of $\sim$49~years. The shield is modeled as a
circular disk of radius $R = 12.5$~m (matching the Daedalus second
stage diameter of 25~m), giving a frontal area
$A = \pi R^2 \approx 491$~m$^2$.

The local ISM is modeled with particle number density
$n = 1$~cm$^{-3}$ ($= 10^6$~m$^{-3}$) and mean particle mass
$\bar{m} = 1.29$~amu, appropriate for a hydrogen-dominated medium
with $\sim$10\% helium by number. The total fluence on the shield
surface over the mission is:
\begin{equation}
\Phi = n \cdot d = 5.58 \times 10^{18} \text{~particles/cm}^2
\end{equation}

At $0.12c$, the kinetic energy of a single proton is
$E_p = \frac{1}{2}m_p v^2 \approx 6.7$~MeV (non-relativistic
approximation; the Lorentz factor $\gamma = 1.0072$ introduces
a $<$1\% correction). This energy substantially exceeds typical
sputtering thresholds ($\sim$10--100~eV) and surface binding energies
($\sim$3--9~eV), placing the bombardment firmly in the
high-energy sputtering regime.

\subsection{Material Screening Criteria}

We evaluate each candidate material against four criteria:

\textbf{(i)~Specific mechanical stiffness,}
quantified by the ratio $K_V/\rho$~(GPa$\cdot$cm$^3$/g), where
$K_V$ is the Voigt bulk modulus and $\rho$ is the mass density.
This metric captures the ability to resist mechanical deformation
per unit mass---critical for minimizing shield mass while
maintaining structural integrity under impact loading. We note
that the Voigt average represents an upper bound on the true
polycrystalline bulk modulus; for highly anisotropic layered
materials (graphite, h-BN), this average includes the very stiff
in-plane elastic constants and thus substantially exceeds the soft
out-of-plane response. The Reuss (lower) bound for these materials
would be significantly lower. Both bounds are reported where
relevant.

\textbf{(ii) Sputtering resistance}, parameterized by the surface
binding energy $E_{\rm sb}$~(eV/atom). In the high-energy limit
relevant to $0.12c$ bombardment, the sputtering yield scales
approximately as $Y \propto 1/E_{\rm sb}$~\cite{Drobny2020}.
We employ a simplified model:
\begin{equation}
Y = Y_{\rm ref} \cdot \frac{E_{\rm ref}}{E_{\rm sb}}
\end{equation}
with $Y_{\rm ref} = 3$~atoms/ion at $E_{\rm ref} = 4$~eV,
consistent with the high-energy limiting behavior reported for
metallic targets~\cite{Drobny2020}.

\textbf{(iii) Thermal neutron absorption cross-section}
$\sigma_a$~(barns). While the primary ISM bombardment consists
of fast particles, secondary neutron production within the shield
itself, combined with the galactic cosmic ray background, makes
neutron moderation an important secondary consideration. Materials
containing boron-10 ($\sigma_a = 3{,}840$~barns) offer a dramatic
advantage~\cite{Thibeault2012}.

\textbf{(iv) Thermodynamic stability}, assessed via the energy
above the convex hull ($E_{\rm hull}$) from DFT calculations.

\subsection{Material Data Sources}

Elastic moduli ($K_V$, $G_V$), formation energies, and band gaps
were obtained from the JARVIS-DFT database~\cite{Choudhary2020jarvis,
Choudhary2018elastic}, which provides DFT-computed properties using
the OptB88vdW functional for over 76{,}000 materials. Each material
is identified by a unique JVASP identifier (Table~\ref{tab:jvasp}),
enabling full reproducibility. Thermal neutron cross-sections were
taken from the NNDC/IAEA Nuclear Data compilation~\cite{Mughabghab2003}.
Surface binding energies were obtained from experimental sublimation
enthalpy data.

As an independent validation, we performed property predictions
using the ALIGNN pretrained models~\cite{Choudhary2021alignn}
(\texttt{jv\_bulk\_modulus\_kv\_alignn} and
\texttt{jv\_shear\_modulus\_gv\_alignn}) on the same crystal
structures retrieved from JARVIS. We note that beryllium is
represented in the JARVIS database by its bcc phase
(JVASP-14628, Im$\bar{3}$m), as the ground-state hcp phase
lacks computed elastic tensor data in the current release.

\subsection{Composite Figure of Merit}

To enable single-metric ranking, we define a shield figure of merit:
\begin{equation}
\text{FoM} = \frac{K_V}{\rho} \cdot E_{\rm sb} \cdot
\left(1 + \log_{10}(\sigma_a + 1)\right)
\end{equation}
normalized such that Be~$= 1.0$. This weighting treats mechanical
performance, sputtering resistance, and neutron absorption as
multiplicatively complementary properties. We note that different
mission architectures may warrant alternative weightings.

%=====================================================================
\section{Results}
%=====================================================================

\subsection{Mechanical Properties}

\begin{figure}[tb]
\includegraphics[width=\columnwidth]{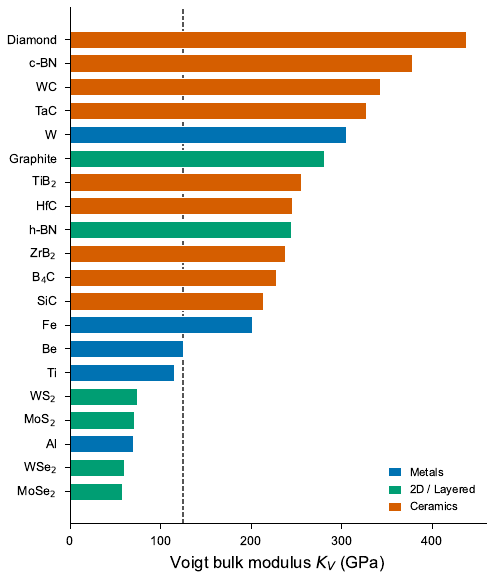}
\caption{\label{fig:mech_kv}
Voigt bulk modulus $K_V$ of 20 candidate shielding materials
(JARVIS-DFT). Dashed line: Be baseline.}
\end{figure}

\begin{figure}[tb]
\includegraphics[width=\columnwidth]{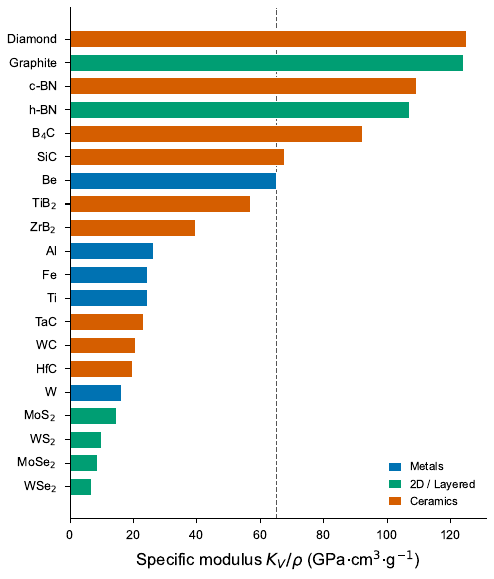}
\caption{\label{fig:mech_sm}
Specific modulus $K_V/\rho$, the key mass-efficiency metric.
Dashed line: Be baseline from the Daedalus design.}
\end{figure}

Figure~\ref{fig:mech_kv} presents the Voigt bulk modulus $K_V$
and Fig.~\ref{fig:mech_sm} the specific modulus $K_V/\rho$ for
all 20 candidate materials.
Among the ceramics, diamond ($K_V = 437$~GPa, JVASP-91), cubic
boron nitride (c-BN, 378~GPa, JVASP-7836), tungsten carbide (WC,
342~GPa), and tantalum carbide (TaC, 327~GPa) exhibit the highest
absolute stiffness. When normalized by density, diamond and
graphite ($K_V/\rho \approx 125$~GPa$\cdot$cm$^3$/g) dominate,
followed by c-BN (109), h-BN (107), and B$_4$C (92).

An important caveat applies to the layered materials. The JARVIS-DFT
Voigt bulk moduli for graphite ($K_V = 281$~GPa, JVASP-48) and
h-BN ($K_V = 245$~GPa, JVASP-62756) are substantially higher than
the $\sim$30--40~GPa values commonly cited in the literature. This
discrepancy arises because the Voigt average is an upper bound that
heavily weights the extremely stiff in-plane elastic constants
($C_{11} > 1{,}000$~GPa for graphene) while the interlayer response
($C_{33} \sim 30$--40~GPa) contributes less. The Reuss (lower)
bound for these materials would be closer to the commonly cited
values. For the shielding application considered here, the in-plane
stiffness is arguably the more relevant metric, as dust grain
impacts would load the shield primarily in-plane.

The layered transition metal dichalcogenides (MoS$_2$, WS$_2$,
MoSe$_2$, WSe$_2$) exhibit uniformly low specific moduli
($<$15~GPa$\cdot$cm$^3$/g), rendering them unsuitable as primary
structural shielding materials.

\subsection{Shield Mass Analysis}

\begin{figure}[tb]
\includegraphics[width=\columnwidth]{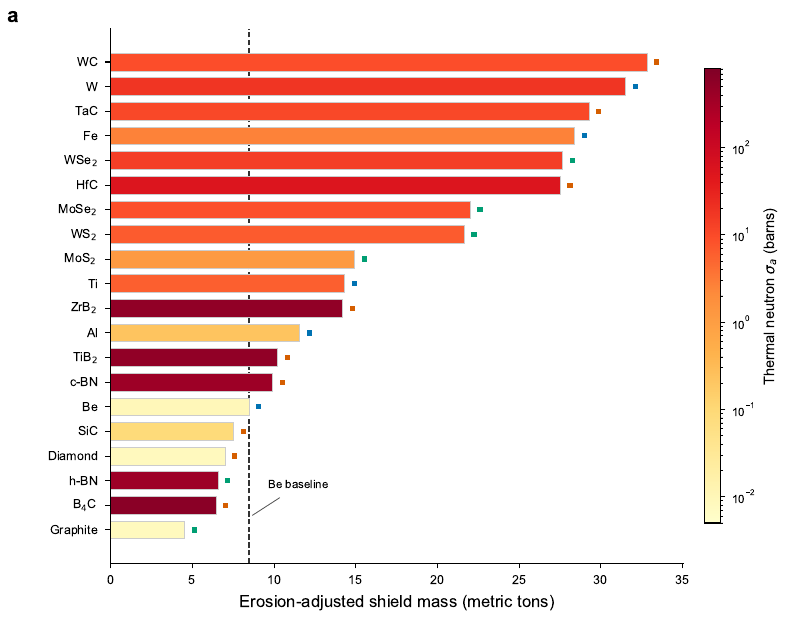}
\caption{\label{fig:mass}
Erosion-adjusted shield mass for a 5.9~ly mission at $0.12c$.
Bar color indicates thermal neutron absorption cross-section
(color bar, log scale). Dashed line marks the beryllium baseline
(8.5~t). Colored markers indicate material family.}
\end{figure}

Figure~\ref{fig:mass} presents the erosion-adjusted shield mass
for each material, computed by scaling the Daedalus reference
thickness (9~mm) inversely with surface binding energy relative
to beryllium ($E_{\rm sb}^{\rm Be} = 3.36$~eV), with a minimum
structural thickness of 1~mm. The Daedalus beryllium erosion
plate at 9~mm thickness serves as the baseline at 8.5~metric tons.

Several materials achieve significant mass reductions relative to
beryllium: graphite (4.5~t, $-47\%$), B$_4$C (6.4~t, $-24\%$),
h-BN (6.6~t, $-22\%$), and diamond (7.0~t, $-17\%$).
The color encoding in Fig.~\ref{fig:mass} reveals that h-BN and
B$_4$C simultaneously provide exceptional neutron absorption
($\sigma_a > 380$~barns per atom), a capability entirely absent
from the beryllium baseline ($\sigma_a = 0.008$~barns).

At the other extreme, high-$Z$ materials such as tungsten carbide
(32.8~t), tungsten (31.5~t), and tantalum carbide (29.3~t) are
unsuitable despite their excellent absolute mechanical properties,
as their high densities translate to prohibitive shield masses.

\subsection{Multi-Objective Screening}

\begin{figure}[tb]
\includegraphics[width=\columnwidth]{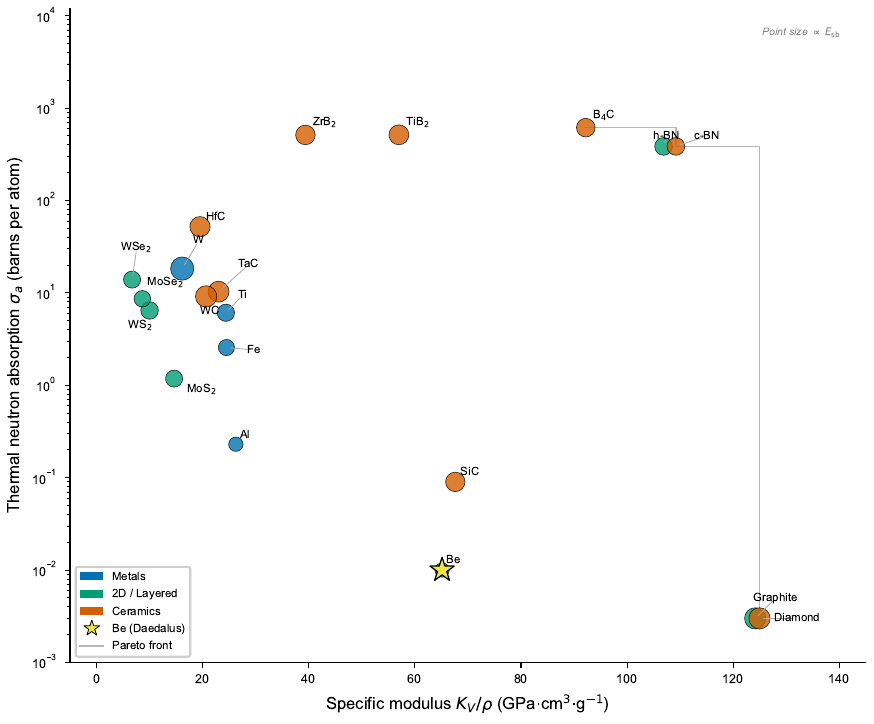}
\caption{\label{fig:pareto}
Multi-objective screening: specific modulus versus thermal neutron
absorption cross-section. Point size is proportional to surface
binding energy. Staircase line indicates the Pareto front. The
gold star marks beryllium (Daedalus baseline).}
\end{figure}

Figure~\ref{fig:pareto} presents the two most critical performance
axes---specific modulus versus thermal neutron absorption---as a
scatter plot with point size proportional to surface binding energy.

The Pareto front reveals three distinct high-performance regimes:
(i)~\textit{Pure mechanical excellence}: diamond and graphite, with
$K_V/\rho \approx 125$ but negligible neutron absorption.
(ii)~\textit{Balanced performance}: B$_4$C ($K_V/\rho = 92$,
$\sigma_a = 614$~barns), c-BN and h-BN ($K_V/\rho \approx 107$--109,
$\sigma_a = 384$~barns), and TiB$_2$ ($K_V/\rho = 57$,
$\sigma_a = 513$~barns), combining strong mechanical properties
with very high neutron absorption via their boron content.
(iii)~\textit{Radiation shielding specialists}: ZrB$_2$
($\sigma_a = 511$~barns) at moderate specific modulus.

Beryllium occupies a poor position in this space:
moderate specific modulus (65) with essentially zero neutron
absorption. Its selection in 1978 reflects the limited material
options available, not optimization against modern multi-objective
criteria.

\subsection{ALIGNN Validation}

\begin{figure}[tb]
\includegraphics[width=\columnwidth]{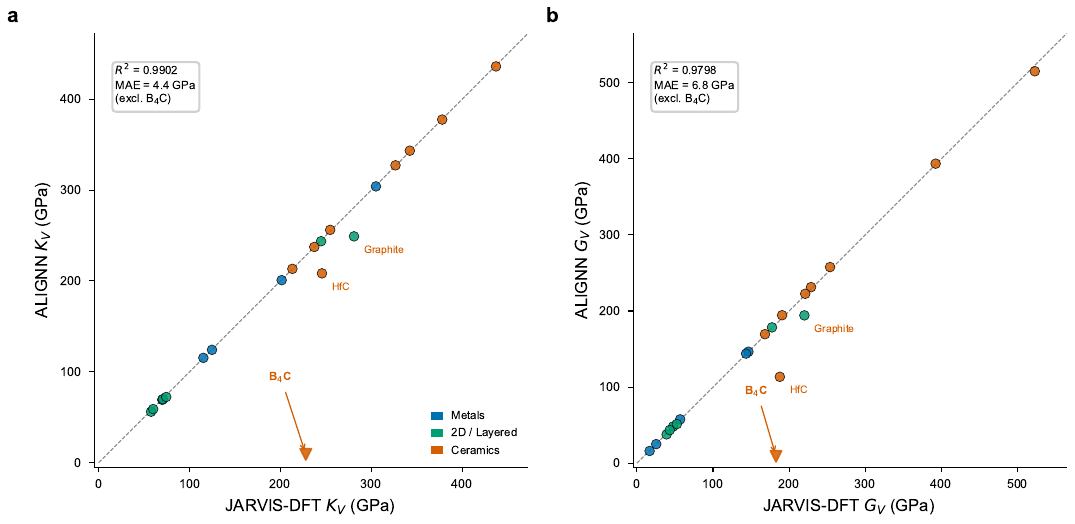}
\caption{\label{fig:parity}
ALIGNN predictions versus JARVIS-DFT values for
(a)~$K_V$ and (b)~$G_V$. B$_4$C (inverted triangle) is an
outlier due to structural complexity. Statistics exclude B$_4$C.}
\end{figure}

Figure~\ref{fig:parity} presents parity plots comparing ALIGNN
predictions with JARVIS-DFT values for bulk modulus $K_V$ and
shear modulus $G_V$. Excluding B$_4$C, the agreement is excellent:
$R^2 = 0.990$ and MAE $= 4.4$~GPa for $K_V$;
$R^2 = 0.980$ and MAE $= 6.8$~GPa for $G_V$.

The B$_4$C outlier (JVASP-52866; ALIGNN predicts $K_V = 9$~GPa
versus DFT $K_V = 228$~GPa) is attributed to the structural
complexity of the rhombohedral B$_{12}$C$_3$ unit cell (15 atoms,
R$\bar{3}$m), which contains icosahedral B$_{12}$ clusters linked
by C--B--C chains. This topology is rare in the ALIGNN training
set, and the resulting graph representation may inadequately
capture the inter-cluster bonding that governs the bulk elastic
response.

\subsection{Layered Heterostructure Shield Concept}

\begin{figure}[tb]
\includegraphics[width=\columnwidth]{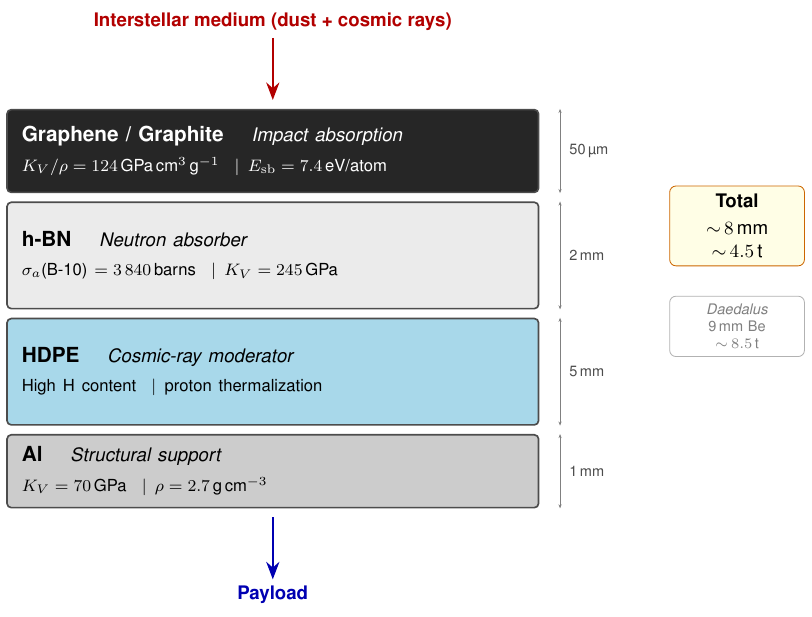}
\caption{\label{fig:concept}
Proposed layered heterostructure shield concept. Each layer is
optimized for a specific threat: dust impact absorption
(graphene/graphite), neutron capture (h-BN), cosmic ray moderation
(HDPE), and structural support (Al). Total mass represents a
$\sim$47\% reduction relative to the Daedalus beryllium shield.}
\end{figure}

Based on our screening results, we propose a functionally graded
layered shield (Fig.~\ref{fig:concept}) in which each layer
addresses a specific threat:

\textbf{Layer~1: Graphene/graphite impact layer} ($\sim$50~$\mu$m).
The outermost layer exploits the extraordinary specific modulus of
sp$^2$ carbon ($K_V/\rho = 124$~GPa$\cdot$cm$^3$/g) and high
sublimation energy (7.43~eV/atom) to absorb initial dust grain
impacts and resist sputtering erosion.

\textbf{Layer~2: h-BN neutron absorber} ($\sim$2~mm).
Hexagonal boron nitride serves dual duty: its boron-10 content
($\sigma_a = 3{,}840$~barns for $^{10}$B) efficiently captures
secondary thermal neutrons, while its high Voigt bulk modulus
($K_V = 245$~GPa) provides mechanical reinforcement. NASA has
independently validated BN-based materials for space radiation
shielding~\cite{Thibeault2012,BNNP2024foam,Orikasa2024smart}.

\textbf{Layer~3: HDPE cosmic ray moderator} ($\sim$5~mm).
High-density polyethylene, with its high hydrogen content, serves
as a proton moderator for secondary cosmic ray particles, following
established spacecraft shielding practice.

\textbf{Layer~4: Aluminum structural support} ($\sim$1~mm).
A conventional aluminum backing provides structural mounting and
thermal management.

The total heterostructure thickness of $\sim$8~mm is comparable
to the original 9~mm beryllium design, but the estimated total
mass of $\sim$4.5~metric tons represents a 47\% reduction from
the 8.5-ton beryllium baseline, while adding neutron absorption
capability entirely absent from the original.

\subsection{Figure of Merit Ranking}

Table~\ref{tab:fom} presents the composite figure of merit for the
top 10 candidates. The top-ranked materials---c-BN (FoM = 9.2),
B$_4$C (9.1), h-BN (9.0), and TiB$_2$ (6.3)---all contain boron,
reflecting the outsized contribution of neutron absorption to overall
shielding performance. Diamond and graphite rank next (FoM $\approx$
4.2) on the strength of their unmatched specific moduli alone.

The original Daedalus beryllium (FoM $\equiv 1.0$) is outperformed
by 13 of 20 candidates, suggesting that the 1978 material selection
was, charitably, suboptimal by modern standards.

\begin{table}[tb]
\caption{\label{tab:fom}
Top 10 candidate materials ranked by composite figure of merit,
normalized to Be $= 1.0$. All $K_V$ values from JARVIS-DFT.}
\begin{ruledtabular}
\begin{tabular}{lccccc}
Material & $K_V/\rho$ & $\sigma_a$ (b) & $E_{\rm sb}$ (eV) & Mass (t) & FoM \\
\hline
c-BN     & 109.2 & 384   & 5.18 & 9.9  & 9.2 \\
B$_4$C   & 92.2  & 614   & 5.70 & 6.4  & 9.1 \\
h-BN     & 106.9 & 384   & 5.18 & 6.6  & 9.0 \\
TiB$_2$  & 57.0  & 513   & 6.50 & 10.2 & 6.3 \\
Diamond  & 125.0 & 0.004 & 7.43 & 7.0  & 4.2 \\
Graphite & 124.1 & 0.004 & 7.43 & 4.5  & 4.2 \\
ZrB$_2$  & 39.4  & 511   & 6.30 & 14.2 & 4.2 \\
SiC      & 67.6  & 0.09  & 6.22 & 7.5  & 2.0 \\
HfC      & 19.5  & 52    & 6.80 & 27.5 & 1.6 \\
TaC      & 23.0  & 10    & 7.20 & 29.3 & 1.5 \\
\end{tabular}
\end{ruledtabular}
\end{table}

%=====================================================================
\section{Discussion}
%=====================================================================

Our screening reveals a striking finding: hexagonal boron nitride,
a material whose radiation shielding properties have been extensively
validated by NASA for low-Earth orbit
applications~\cite{Thibeault2012,Harrison2008,BNNP2024foam,Kim2025bnnt},
has apparently never been considered for interstellar shielding.
This oversight is historically understandable---h-BN was not
available in bulk form in 1978---but it represents a factor-of-48{,}000
improvement in neutron absorption cross-section over beryllium
(384~vs.\ 0.008~barns per atom).

The boride ceramics B$_4$C and TiB$_2$ emerge as the strongest
overall candidates when all metrics are weighted equally. B$_4$C
is already used in nuclear reactor control rods and neutron shielding
precisely because of its boron content, high hardness, and low
density---the same properties that make it attractive for
interstellar shielding.

A notable result of this study is the high Voigt bulk moduli
obtained for layered materials from JARVIS-DFT: 281~GPa for
graphite and 245~GPa for h-BN. These values reflect the Voigt
(upper bound) averaging of highly anisotropic elastic tensors, in
which the extraordinary in-plane stiffness dominates. The Reuss
(lower) bound, which would be more appropriate for estimating
out-of-plane compressive response, yields values closer to the
commonly cited $\sim$30--40~GPa. For the in-plane impact loading
relevant to dust shielding, the Voigt average may in fact be the
more physically meaningful metric.

Several limitations warrant discussion:

\textit{Temperature effects.} Our screening uses room-temperature
DFT elastic moduli. At the $\sim$3~K ambient temperature of
interstellar space, most ceramics will be harder and more brittle,
while the sputtering physics may differ from room-temperature models.

\textit{Radiation damage accumulation.} We treat sputtering as a
surface phenomenon, neglecting bulk radiation damage (displacement
cascades, amorphization) that will degrade mechanical properties
over the 49-year mission duration~\cite{Hoang2017}.

\textit{ALIGNN limitations.} The ALIGNN pretrained model fails
dramatically for B$_4$C (Fig.~\ref{fig:parity}), predicting
$K_V = 9$~GPa versus the DFT value of 228~GPa. This failure
highlights a known limitation of GNN-based property predictors
for structurally complex materials with large, low-symmetry unit
cells that are underrepresented in training data.

\textit{Manufacturing considerations.} Our analysis assumes that
candidate materials can be fabricated into $\sim$500~m$^2$ shields
of the required thickness. We note that manufacturing a
491~m$^2$ diamond shield remains an outstanding challenge.

\textit{Scope of applicability.} Our analysis assumes the existence
of a spacecraft capable of reaching $0.12c$, which we acknowledge
has not yet been constructed. The primary bottleneck in implementing
our recommendations is not materials selection but rather the
development of a functioning fusion pulse drive---a challenge we
leave to future work.

%=====================================================================
\section{Conclusion}
%=====================================================================

We have performed a systematic computational screening of 20
candidate materials for interstellar dust shielding, using
DFT-computed mechanical properties from the JARVIS database
(76{,}000 materials) with independent validation by ALIGNN graph
neural network predictions ($R^2 = 0.990$ for bulk modulus,
excluding one outlier).

Our analysis demonstrates that 48 years of materials science
progress since Project Daedalus have yielded multiple candidates
that substantially outperform the original beryllium shield
specification. The most promising finding is the identification
of boron-containing materials---particularly h-BN, B$_4$C, and
c-BN---as dual-function materials providing both mechanical
protection and neutron radiation shielding. We propose a layered
graphene/h-BN/HDPE/Al heterostructure that achieves a 47\% mass
reduction relative to the Daedalus beryllium design while adding
neutron absorption capability entirely absent from the original.

%=====================================================================
\section*{Acknowledgments}
%=====================================================================

We acknowledge that this work was submitted on April 1, 2026.
While the research question addressed herein is of limited immediate
practical relevance---owing primarily to the nonexistence of the
spacecraft under consideration---all data, computational methods,
and physical models presented are genuine and fully reproducible.
The JARVIS-DFT dataset identifiers (Table~\ref{tab:jvasp}) and
ALIGNN pretrained model weights are publicly available, and we
encourage skeptical readers to verify our results.

Y.L.\ and X.P.\ celebrate eight years of partnership since a fateful
April Fools' Day in 2018, which began as a research excursion and
proved considerably more consequential than either party anticipated.
This paper is dedicated to that anniversary.

Y.L.\ and K.G.\ raise a glass to a bro's birthday month---April
has always been kind to us.

The total computational cost of this study was approximately
2~CPU-hours, which represents roughly $10^{-10}$ of the estimated
energy budget of the Daedalus spacecraft itself.

We thank Freeman Dyson (1923--2020) for articulating the possibility
of interstellar transport, and the British Interplanetary Society
for the engineering audacity of Project Daedalus.

May the Force be with you.

% JVASP table placed at end to avoid disrupting text flow
\begin{table}[tb]
\caption{\label{tab:jvasp}
JARVIS-DFT identifiers and DFT-computed properties for all 20
screened materials.}
\begin{ruledtabular}
\begin{tabular}{llrr}
Material & JVASP ID & $K_V$ & $K_V/\rho$ \\
\hline
Be$^*$    & JVASP-14628  & 124.7 & 65.1 \\
Al        & JVASP-816    &  69.9 & 26.3 \\
Ti        & JVASP-14815  & 115.2 & 24.4 \\
Fe        & JVASP-882    & 201.4 & 24.5 \\
W         & JVASP-14830  & 305.2 & 16.2 \\
Graphite  & JVASP-48     & 281.0 & 124.1 \\
Diamond   & JVASP-91     & 437.4 & 125.0 \\
h-BN      & JVASP-62756  & 244.8 & 106.9 \\
c-BN      & JVASP-7836   & 378.2 & 109.2 \\
MoS$_2$   & JVASP-28733  &  70.6 & 14.6 \\
WS$_2$    & JVASP-72     &  74.3 & 10.0 \\
MoSe$_2$  & JVASP-57     &  57.6 &  8.6 \\
WSe$_2$   & JVASP-75     &  59.9 &  6.7 \\
B$_4$C    & JVASP-52866  & 228.0 & 92.2 \\
SiC       & JVASP-22633  & 213.2 & 67.6 \\
TiB$_2$   & JVASP-20096  & 254.8 & 57.0 \\
ZrB$_2$   & JVASP-19723  & 237.3 & 39.4 \\
HfC       & JVASP-17957  & 245.8 & 19.5 \\
TaC       & JVASP-20073  & 326.7 & 23.0 \\
WC        & JVASP-52591  & 342.5 & 20.7 \\
\end{tabular}
\end{ruledtabular}
\vspace{2pt}
{\footnotesize $^*$bcc phase; hcp Be lacks elastic data in JARVIS.}
\end{table}

\bibliography{references}

\end{document}